\documentclass[aip,apl,reprint]{revtex4-2}

\usepackage{graphicx}

\usepackage[utf8]{inputenc}
\usepackage[T1]{fontenc}
\usepackage{mathptmx}
\usepackage{etoolbox}

\usepackage{tabularray}
\UseTblrLibrary{booktabs}

\begin{document}

\title{Site-controlled quantum dot arrays edge-coupled to integrated silicon nitride waveguides and devices}

\author{John O'Hara}
\email{john.ohara@tyndall.ie}
\affiliation{Tyndall National Institute, Lee Maltings, Cork, Ireland}
\author{Nicola Maraviglia}
\affiliation{Tyndall National Institute, Lee Maltings, Cork, Ireland}
\author{Mack Johnson}
\affiliation{Tyndall National Institute, Lee Maltings, Cork, Ireland}
\author{Jesper Håkansson}
\affiliation{Tyndall National Institute, Lee Maltings, Cork, Ireland}
\author{Salvador Medina}
\affiliation{Tyndall National Institute, Lee Maltings, Cork, Ireland}
\author{Gediminas Juska}
\affiliation{Tyndall National Institute, Lee Maltings, Cork, Ireland}
\author{Luca Colavecchi}
\affiliation{Tyndall National Institute, Lee Maltings, Cork, Ireland}
\author{Frank H. Peters}
\affiliation{Tyndall National Institute, Lee Maltings, Cork, Ireland}
\affiliation{School of Physics, University College Cork, College Road, Cork, Ireland}
\author{Brian Corbett}
\affiliation{Tyndall National Institute, Lee Maltings, Cork, Ireland}
\author{Emanuele Pelucchi}
\affiliation{Tyndall National Institute, Lee Maltings, Cork, Ireland}

\date{\today}

\begin{abstract}
The scalability of quantum photonic integrated circuits opens the path towards large-scale quantum computing and communication. To date, this scalability has been limited by the stochastic nature of the quantum light sources. Moreover, hybrid integration of different platforms will likely be necessary to combine state-of-the-art devices into a functioning architecture. Here, we demonstrate the active alignment and edge-coupling of arrays of ten site-controlled gallium arsenide quantum dots to an array of ten silicon nitride single-mode waveguides, at cryogenic temperatures. The coupling is facilitated by the fabrication of nanopillars, deterministically self-aligned around each quantum dot, leading to a high-yield and regular array of single-photon sources. An on-chip beamsplitter verifies the triggered emission of single photons into the silicon nitride chip. The low inhomogeneous broadening of the ensemble enables us to observe the spectral overlap of adjacent site-controlled emitters. Across the array of waveguides, the signal collected from each coupled quantum dot is consistently and reproducibly 0.17 relative to the free-space collection from the very same single-photon source. Comparing measurement with waveguide simulations, we infer that absolute coupling efficiencies of $\approx 5 \%$ are currently obtained between our quantum dots and the waveguides.
\end{abstract}

\maketitle

\section{Introduction}

The promise of light as the medium in which to realize large-scale quantum information processing rests on the unique properties of the photon \cite{walmsley_light_2023, romero_photonic_2025} and the theoretical scalability of quantum photonics \cite{pelucchi_potential_2022, labonte_integrated_2024, psiquantum_team_manufacturable_2025}. In particular, the latter arises from the photon's robustness against environmental decoherence, and the ability to manufacture dense photonic integrated circuits (PICs). However, the highest quality single-photon sources (SPSs) remain probabilistic, either in growth position -- as with randomly nucleated quantum dots (QDs) \cite{gurioli_droplet_2019, da_silva_gaas_2021} and colour centres \cite{katsumi_recent_2025}, or in photon number -- as with spontaneous parametric down-conversion and four-wave mixing \cite{signorini_-chip_2020}. Thus the ability to create dense ordered arrays of deterministic quantum emitters and combine them with integrated circuits has not yet been demonstrated.

Most schemes for photonic quantum information processing involve a path through the distinct steps of deterministic light generation, manipulation, and detection \cite{kim_hybrid_2020}. Currently, the best performing options for the various quantum PIC components are to be found across a range of platforms \cite{pelucchi_potential_2022, labonte_integrated_2024}, and all of these face their own challenges to become a high-yield technology. For these reasons, a modular hybrid approach \cite{elshaari_hybrid_2020, kaur_hybrid_2021}, wherein the fundamental operations remain confined to their respective functionally separate chips, and are combined by leveraging advances in photonic coupling and packaging, has become a common approach to scalability \cite{marchetti_coupling_2019}.

In early work on QD integration, devices were fabricated using high density QD samples. This increased the chance of having a QD well-positioned within the device, and relied on the large inhomogeneous broadening of the ensemble in order to be able to filter out the extraneous emitters. This approach is still the groundwork for proof-of-principle demonstrations \cite{dusanowski_-chip_2023, hornung_highly_2024}, which indeed often exploit the random variations in the dot-field characteristics. Nevertheless, deterministic integration of such QDs with photonic circuits is then only possible by pre-selecting and registering QDs, and tailoring the circuit design to the source positions \cite{ollivier_reproducibility_2020, schnauber_spectral_2021, li_scalable_2023, papon_independent_2023}, or by one-by-one transfer of sources to the input waveguides \cite{yeung_-chip_2023}. Neither of these approaches are particularly scalable to many emitters and/or complex circuits. Ultimately, scalable \textit{and} deterministic integration will require site-control of the emitters, with a circuits designed around known and reproducible emitter positions \cite{jamil_-chip_2014}, as well as a method to tune the emitters into mutual indistinguishability, if necessary.

Site-control of SPSs has been achieved by a variety of methods, including patterning of growth substrates with nanoholes or pyramidal recesses \cite{skiba-szymanska_narrow_2011, jons_triggered_2013}, pre-strained substrates \cite{straus_resonance_2017}, growth on nanostructure arrays \cite{dalacu_deterministic_2010, lagoudakis_ultrafast_2016, zhang_-chip_2022}, as well as laser-guided growth \cite{wang_precise_2020}. So far, integration of multiple site-controlled QDs with a PIC has not been demonstrated. Here we show that it is possible to actively align and couple an array of site-controlled quantum dots in self-aligned nanopillars \cite{juska_self-aligned_2025} to an array of single-mode waveguides of matching pitch, and to other integrated photonic devices. We demonstrate the low inhomogeneous broadening and high single-photon purity of the QDs. Lastly, we discuss the advantages, current limitations, and future directions of this approach, and compare it with other integration strategies.

\section{Results}

\subsection{Devices}

Our light sources consisted of arrays of single site-controlled GaAs QDs, with neutral excitonic emission around 778 nm. All samples were grown via metal-organic vapour-phase epitaxy in a hexagonal array of lithographically defined pyramidal recesses \cite{juska_towards_2013}, in this case with a 10 {\textmu}m pitch. The optical lithography together with the wet etch process resulted in a standard deviation of each recess position of less than 50 nm with respect to the ideal local lattice node, as confirmed by SEM measurement. During growth, the QDs (which have a lateral dimension of $\approx 15$ nm) naturally align with nanometre precision in the centre of each recess \cite{pelucchi_self-ordered_2018, dimastrodonato_self-limiting_2012}. The QDs were grown between GaAs/AlAs superlattices \cite{ranjbar_jahromi_optical_2021} and Fe-doped AlGaAs barriers.

\begin{figure}
\centering
\includegraphics[width=\columnwidth]{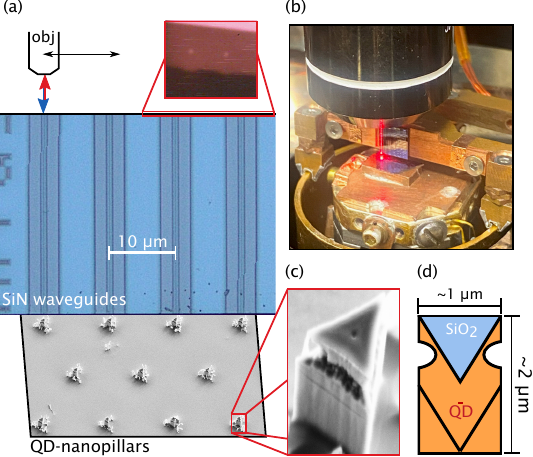}
\caption{\label{fig:expmtl} Experimental arrangement and sample details. (a) The QD-nanopillar array positioned orthogonally underneath a SiN single-mode waveguide array, with SEM and optical images of the chips, respectively. Both the SiN waveguides and the QD-pillars are separated with 10 {\textmu}m pitch. The top inset shows an optical image of the waveguide facets. (b) Photo of the experimental arrangement used to achieve the edge-coupling, with a red laser coupled through a SiN waveguide down to the QD chip. (c) A close-up of one of the fabricated nanopillars. (d) A schematic of the nanopillar, showing the lateral etching. For clarity, the carrier confinement layers immediately above and below the QD have been omitted.}
\end{figure}

In order to improve the far-field emission profile and increase the coupling to waveguides, photonic nanopillars were shaped around the QDs through post-growth processing \cite{juska_self-aligned_2025}. This procedure automatically aligns the centre of the triangular cross-section pillars with the QD. The nanostructures had a lateral dimension of around 1.2 {\textmu}m, and an additional lateral etching occurred due to the epitaxial layers and the etching recipe, as shown in figure \ref{fig:expmtl}(c).

For the PIC we utilized SiN waveguides and devices in SiO$_2$ cladding on a 700 {\textmu}m thick Si substrate. The single transverse-electric (TE) and transverse-magnetic mode waveguides, in order to operate at 780 nm, were defined with 400 nm width in a 300 nm thick layer of SiN, with 2 {\textmu}m and 3 {\textmu}m of SiO$_2$ above and below respectively. These waveguides tapered down to 200 nm width at the facet. The designed effective mode area for the 780 nm TE-mode was 0.21 {\textmu}m$^2$ in the main waveguide, and expanded to 0.33 {\textmu}m$^2$ at the facet. The chips were fabricated at a commercial foundry (Cornerstone), where the features were defined by optical stepper lithography, giving a centre-centre uncertainty in waveguide position of $<25$ nm. Subsequently, the facets were commercially polished.

The PICs consisted of a range of nanophotonic devices and gates. The structures relevant to this paper include the simple straight waveguides, the array of ten waveguides, and the 2x2 multi-mode interferometer (MMI) that functioned as a beamsplitter. On one side of the PIC, the waveguide facets for each multi-port device were arranged with 10 {\textmu}m separation in order to match the QD array. Since, at present, it is necessary to retrieve the light from the PIC in order to complete the measurements, all of the devices had output facets on the opposite edge of the chip. In this case there was no strict requirement on the spacing, but the facets were placed further apart to avoid cross-talk and background light. Thus, the waveguide arrays fanned out to have 20 {\textmu}m separation at the outputs, while the MMI outputs were 50 {\textmu}m apart.

\subsection{Alignment procedure}

\begin{figure*}
\centering
\includegraphics[width=\linewidth]{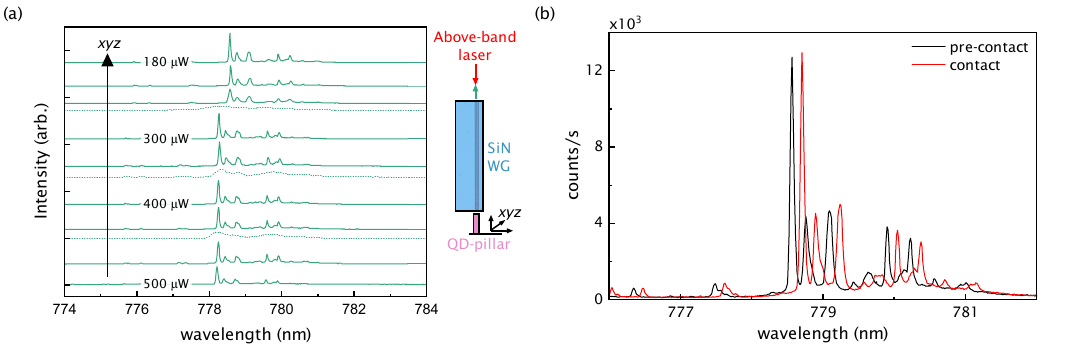}
\caption{\label{fig:align} Active alignment of a QD-pillar to a single-mode waveguide. (a) Final stages of alignment. The spectra are offset but shown on the same scale. The excitation power was set to give the maximum signal and the QD-pillar was aligned in $xyz$, and these steps were iteratively optimized. After the signal was close to maximum, further improvement was noticeable by a several-fold reduction in the excitation powers, which are indicated in the figure. Also shown (dotted lines) are spectra taken during over-contact, where the spectrum was significantly broadened -- this was then reversed by increasing the $z$ distance. (b) In the final stages of chip-chip alignment, the QD chip could come into contact with the PIC, causing a small redshift due to strain. Small redshifts were generally reversible by increasing the separation.}
\end{figure*}

In our approach the PIC chip was fixed to a copper mount such that the bottom facets were suspended, while the QD chip was mounted orthogonally below this on a separate copper mount on an \textit{xyz} piezoelectric stage, and thus could be moved into proximity/contact with the photonic chip with sub-{\textmu}m precision. These were both actively cooled within a closed-cycle helium cryostat (and probe station), shown in figure \ref{fig:expmtl}(b), and optically accessible through a top window. To prevent heating of the QD sample when the PIC was in close proximity, we ensured an efficient thermal contact between the PIC and the inner radiation shield of the cryostat (nominally 12-13 K), see SI for details. 

Before \textit{xyz} alignment of the chips, it was necessary to rotationally align the chips such that the facet plane of the photonic chip was parallel to the surface of the QD chip, and also that the edge of the PIC was aligned with one basis vector of the hexagonal QD array. Both of these alignments were performed manually at room temperature, as detailed in the SI. Assuming one waveguide was optimally aligned, the uncertainty of the angular alignment procedure ($<1$ mrad) corresponded to a misalignment of the adjacent waveguide of $<20$ nm, less than the QD or waveguide positional variation expected from the lithography. See below and the SI for further discussion on alignment tolerances.

Due to the fact that the QD chip and its $xyz$ stage may cool at different rates to the PIC and its mount, the cooling was initiated with the chips apart. It was observed that there was a small systematic rotation between the room temperature position of the PIC (with respect to the QD chip) and its final position at low temperature, and so two of the PIC angles were intentionally slightly offset when prepared at room temperature so that the PIC was optimally aligned with the QD sample when at $\approx 10$ K.

Once the chips were cold we began the active alignment using the QD signal. The quantum dots were excited with a continuous-wave above-band laser (636 nm), coupled through one of the top facets of the PIC, and the QD signal was collected from the same facet, as shown in figure \ref{fig:expmtl}(a) and (b). It was possible to observe the signal from the QDs while still at a relatively large distance from contact (chip-chip separation $\Delta z\geq 100$ {\textmu}m), and so the $xyz$ alignment could be performed with the QD signal alone -- firstly via a multimode waveguide (also of 300 nm nominal thickness, but $\geq 10$ {\textmu}m in width), and subsequently with the single-mode waveguides. Figure \ref{fig:align}(a) shows the final stages of alignment of a QD signal. Over the course of the alignment procedure, the signal increased; however, in the final stages, the effect of improved alignment was mainly in reduced excitation power, rather than increased count rate.

Figure \ref{fig:align}(a) also shows that the spectrum sometimes redshifted abruptly with respect to the initial condition. This suggests that the pillars had contacted the PIC and been compressively strained along their long axis, and that the effect of this was to redshift the QD emission. The optimal coupling was generally observed under a small amount of redshift; however, the improvement from the not-in-contact condition was minimal (see figure \ref{fig:align}(b)). At higher levels of strain (more forceful contact) the spectrum redshifted further before broadening (several such events can be seen in figure \ref{fig:align}(a)). In a small minority of cases, an unusually large shift and/or a change in the average charge state was observed. The effect of this strain was mostly reversed upon disconnection (see also figure \ref{fig:array-comp}).

We verified that the approach to coupling was mechanically and thermally stable. After aligning the QD array to the waveguide chip, the signal was recorded at various intervals over a period of 41 hours (see SI). Without any chip-chip realignment, the signal was stable. This indicates both that the chip-chip alignment was stable and that the QDs were not negatively affected by the (near-)contact.

\subsection{On-chip photon correlation measurements}

\begin{figure}
\centering
\includegraphics[width=\columnwidth]{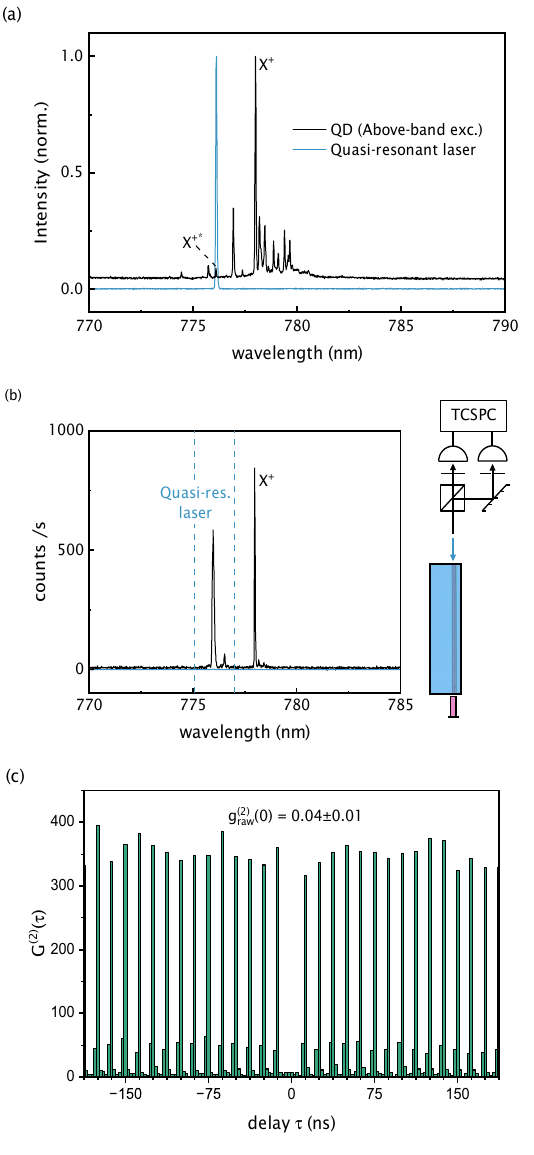}
\caption{\label{fig:corr-offchip} Waveguide-coupled single-photon measurement. (a) Normalized spectra of QD emission obtained through the single-mode waveguide under above-band excitation (black). The quasi-resonant laser (blue) was tuned to one of the hot trion states $X^{+*}$, and subsequently filtered with notch filter(s). (b) Under this excitation the $X^+$ line was selectively excited. TCSPC: time-correlated single-photon counter (c) Raw autocorrelation measurement of the $X^+$ line of a QD-pillar when excited quasi-resonantly through the hot trion $X^{+*}$ state, obtained through a waveguide and an off-chip beamsplitter.}
\end{figure}

To verify that single QDs were coupled to the chip waveguides, a second-order correlation measurement ($G^{(2)}(\tau)$) was recorded for a single QD line. The positive trion state was selectively excited through one of its excited "hot-trion" states (having the additional hole in a higher energy level) using a pulsed Ti:Sapphire laser ($\approx 777$ nm, 80 MHz repetition, shaped to around 10 ps duration) which was coupled into the top facet of the waveguide. Hereafter this is referred to as quasi-resonant excitation. This arrangement is shown schematically in figure \ref{fig:corr-offchip}(b). After collecting the light out of the chip, notch filters suppressed the excitation laser, and an off-chip beamsplitter separated the photons into two paths, which each passed through a monochromator towards single-photon detectors. The zero-delay correlation value, $g^{(2)}_{\text{raw}}(0) = 0.04 \pm 0.01$, confirmed that the single-photon purity was high.

\begin{figure*}
\centering
\includegraphics[width=\linewidth]{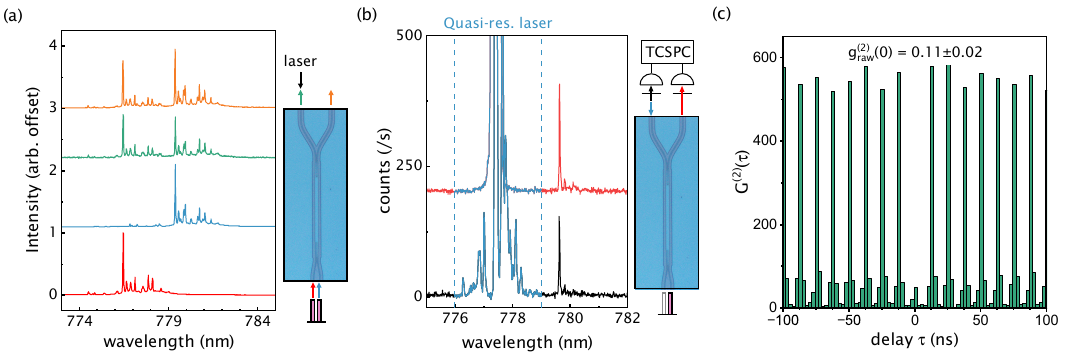}
\caption{\label{fig:corr-onchip} Single-photon verification using the on-chip beamsplitter. (a) Normalized (and effectively background-subtracted) spectra of two adjacent QDs taken directly (red and blue), and through the two output ports of the MMI (green and orange) while exciting both QDs non-resonantly through one port with an above-band laser. (b) Spectra of the QD emission obtained through the two output ports of the MMI when exciting through the hot-trion $X^{+*}$ state. (c) Raw $G^{(2)}(\tau)$ measurement. Correlation of the $X^+$ line from the two output ports of the MMI.}
\end{figure*}

Subsequently, the on-chip MMI beamsplitters were employed. As shown in figure \ref{fig:corr-onchip}(a), the input (QD-side) ports of the MMI were separated with the same pitch as the QD array, and thus two adjacent QD-pillars are coupled and combined by the device into the two output ports. When exciting non-resonantly, the multiple lines that were seen from two adjacent QDs were in general spectrally overlapping (due to the high uniformity of QD sizes discussed below, a general feature of these quantum dots \cite{pelucchi_semiconductor_2012}), and so for this particular demonstration of the device, we had to find two unusually well spectrally-separated nearest neighbour QDs. When this same configuration was excited with the quasi-resonant excitation matched to one of the hot trion transitions, as before, we observed only one dominant $X^+$ transition from one QD-pillar from both MMI output ports. This is shown in figure \ref{fig:corr-onchip}(b). Correlating these signals, we obtained $g^{(2)}_{\text{raw}}(0) = 0.11 \pm 0.02$, which again confirmed the high purity of the single-photon source (for clarity, only 200 ns are shown, see SI for longer range), and also the successful on-chip beam-splitting ($T:R = 0.54:0.46$). As a reference for the QD sample performance, a $g^{(2)}_{\text{raw}}(0)$ value of $0.010 \pm 0.003$ (SI) was obtained through a direct-to-objective measurement of a different QD-pillar. The increased coincidence counts at zero delay in the PIC-coupled cases were mostly due to an increased background light (due to higher input excitation powers).

\subsection{Single-mode waveguide array coupling}

\begin{figure*}
\centering
\includegraphics[width=\linewidth]{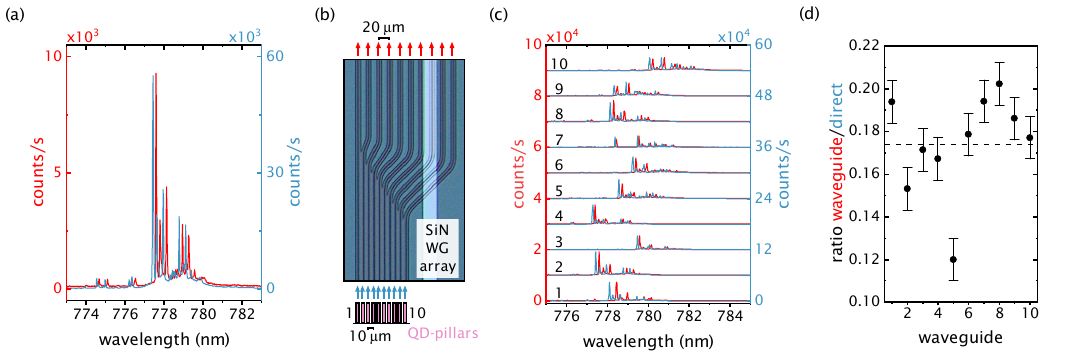}
\caption{\label{fig:array-comp} Coupling to the waveguide array. (a) Comparison of the spectrum of a single QD obtained through a SiN waveguide (red) and directly via an objective lens (blue), at 13 K. (b) Under the same chip-chip alignment, we obtained 10 signals from 10 sequential QDs through the single-mode waveguide array. (c) Spectra of 10 sequential QDs obtained through a SiN single-mode waveguide array (red) and directly via an objective lens (blue). (d) The ratio of the integrated signals observed in (c), with uncertainties due to the peak fitting procedure.}
\end{figure*}

Next we performed the coupling of ten QDs to a ten-waveguide array. After aligning a single QD to a single waveguide of the array, we scanned the laser spot across the array facets, which at the top fanned out to 20 {\textmu}m separation. Without any chip-chip realignment, we observed the signal of every subsequent QD in every subsequent single-mode waveguide, as sketched in figure \ref{fig:array-comp}(b).

As noted above, and as can also be seen in figure \ref{fig:array-comp}(c), the spectrum observed through the waveguides could be slightly redshifted with respect to that observed directly -- between 50 and 300 pm depending on the QD-pillar. For the data reported in figure \ref{fig:array-comp}, the direct-to-objective measurements were obtained after the waveguide array measurement, so in fact we see here the reversal of the contact-induced redshift.

Comparing the light emission measured through the waveguide array, to that from the same pillars measured directly with the objective lens (numerical aperture 0.42), we saw that the detected signal variations arose only from the QD-nanostructures themselves, with the level of pillar-waveguide coupling approximately equal from waveguide 1 to waveguide 10, as shown in figure \ref{fig:array-comp}(c) and (d). Indeed, all ten waveguide spectra were taken at the same excitation power. Thus, after aligning one QD, the entire QD array was successfully aligned to the waveguide array, with the waveguide-coupled signal $0.17 \pm 0.02$ of that measured directly.

\subsection{Absolute QD-waveguide coupling efficiency}

In order to understand the absolute coupling efficiencies between the QDs and the PIC waveguide modes, far-field radiation simulations were performed of the waveguide-objective system and compared with the quasi-resonant excitation experiment. The calculations accounted for the expected propagation losses of the 6 mm waveguide (assumed to be 1 dB/cm), and the transmission and emission profile from the waveguide facet towards the objective. The preparation efficiency of the 80 MHz quasi-resonant excitation was calculated based on a comparison with the above-band excitation, in terms of count rates, spectral profiles, and measured multi-photon contributions. The tabletop efficiency was measured at the QD wavelength 780 nm. Combining these with the observed quasi-resonant waveguide-coupled count rates for the QD-pillar in figure \ref{fig:corr-offchip}, we obtained an estimated QD-waveguide coupling efficiency of $0.05 \pm 0.02$. This value combines the efficiency of the QD dipole to nanopillar coupling (in the direction of the PIC), and the efficiency of the pillar-waveguide coupling. Here the uncertainty is mostly arising from the uncertainty of the preparation efficiency -- see SI for more details.

Across the array, it was observed that the pillar-waveguide coupling was approximately constant, but the absolute count rates varied due to intensity or spectral variations in the QD-pillar sources. Although pulsed-excitation single-photon measurements were not performed systematically across the array, the waveguide-coupled count rates for the QD-pillar of figure \ref{fig:corr-offchip} were very close to the mean value observed during the array measurements, under equivalent conditions. Thus the above estimation of the absolute QD-waveguide coupling is also representative of the mean value obtained across the array in figure \ref{fig:array-comp}. 

\subsection{Coupling simulations and tolerance} \label{simtol}

In order to gain a better understanding of the tolerances of our scheme, and to reveal potential paths to improve the coupling efficiency, we next performed finite-difference time-domain simulations of the QD-to-waveguide and QD-to-objective coupling, as mediated by the nanopillar.

Because the pillars presented an irregular side etch, as shown in figure \ref{fig:expmtl}(c), we adopted a simplified geometry whereby the lateral etch is described by three symmetrical cylindrical grooves. Initially it was assumed that this etch was approximately at the same vertical level as the QD. This geometry produced a collimated far-field and a four to five-fold increase in the collection efficiency when coupling directly to the objective lens, when compared with straight-sided pillars (see SI for the simulation results). Indeed, the fabricated pillars were several times brighter than most previously measured straight-sided pillars with similar lateral dimensions. 

Based on the experimentally observed count rate of the positive trion line under above-band 80 MHz excitation, the measured coupling from QD to objective was calculated to be $0.07 \pm 0.01$. The corresponding value obtained from the simulated geometry was 0.13. For optimal alignment in $xyz$, the same geometry gave a value for the QD-waveguide coupling of 0.11, where $0.05 \pm 0.02$ was observed experimentally. The discrepancy between experiment and modelling are similar in both configurations. Therefore, it may arise from additional unknown losses (e.g. non-radiative states) in the real system, or from imperfect photonic simulation of the pillar.

A comparison of the real QD position, estimated from the growth recipe, with the lateral etch height, obtained from SEM imaging, indicated that the etch was in fact about 500 nm higher than the QD, as shown schematically in figure \ref{fig:expmtl}(d). We proceeded by performing additional parameter sweeps (across four geometrical degrees of freedom: pillar lateral size, QD vertical placement, cylindrical etch vertical placement, and radius of the etched cylinders) to look for an effective geometry that reproduced the experimentally observed emission properties while better matching the SEM images. Of these sets of parameters, none were found to produce a coupling into the objective $\geq 0.044$, leading to the conclusion that either our simplified geometric model was not fully representative of the fabricated pillar, or else the true QD position was indeed closer to the lateral etch than that inferred from the growth recipe and SEM imaging. 

In any case, the pillar geometries (both free and SEM-restricted) that optimized the coupling into the objective showed a saturation trend of the QD-to-waveguide coupling at a chip-chip separation $\Delta z \lesssim 0.5$ {\textmu}m, consistent with experimental observations. Furthermore, the simulation model for the excitation of the nanopillar (which takes into account the pump beam coupled-into and emitted from the PIC towards the nanopillar) confirmed that the QD excitation efficiency continues to increase for smaller chip-chip separations (down to $\Delta z \lesssim 0.1$ {\textmu}m, SI), in agreement with what was observed in the experiments (figure \ref{fig:align}(a)). This, along with the observed wavelength shifting of the emission, suggests that a small $\Delta z$ consistent with close to maximal coupling was obtained. The simulations also showed that the tolerance of the system to $xy$ misalignment was high compared to the minimum piezoelectric stage step size, and experimentally, there was no indication of a cumulative misalignment across the array. The simulations and the experimental observations therefore indicate that the coupling was performed with near-optimal $xyz$-alignment.

\begin{figure}
\centering
\includegraphics[width=\columnwidth]{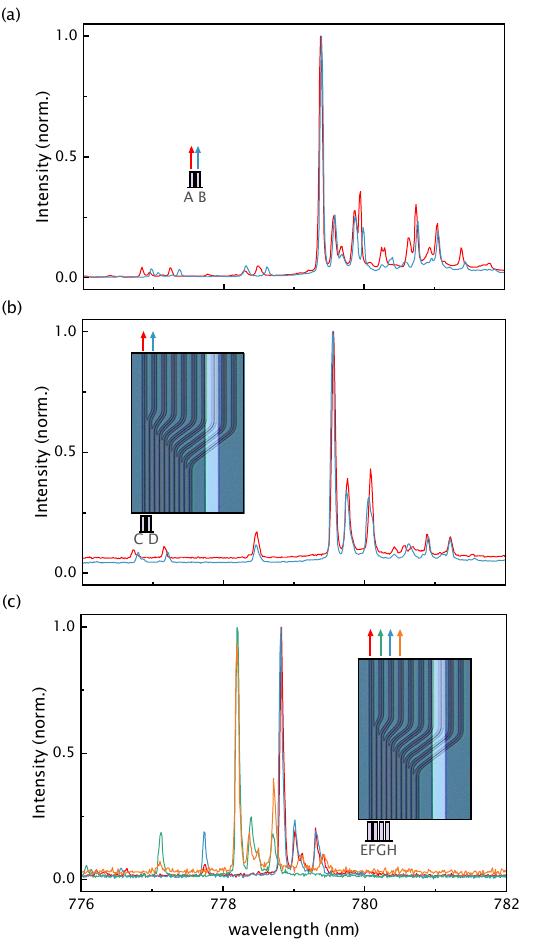}
\caption{\label{fig:unif} Spectral uniformity of the emitter array. All 8 spectra are from different quantum QDs (labelled A to H). Here, even with a low density site-controlled QD array, significant spectral overlap was achieved with adjacent or nearby QDs. (a) Direct spectra from two adjacent QDs. (b) Spectra from two different and adjacent QDs, simultaneously coupled to two adjacent waveguides of the PIC array. (c) Another 4 sequential QDs, which contain two overlapping pairs.}
\end{figure}

\setlength{\heavyrulewidth}{1.5pt}
\begin{table*}[!t]
    \centering
    \resizebox{\linewidth}{!}{
    \begin{booktabs}{
     colspec = {lccccc},
     cell{1}{1} = {c=2}{c},
     cell{2,4}{1} = {r=2}{c},
     cell{6}{1} = {r=6}{c}}
    \toprule
        \textbf{Reference} & ~ & \cite{ellis_independent_2018} & \cite{pfister_telecom_2025} & \cite{wang_large-scale_2025} & \textit{this work} \\ \midrule 
        \textbf{Single-photon source} & Wavelength (nm) & 890-950 & 1530 & 920 & 780 \\
        ~ & Site-controlled & N & N & N & Y \\ \cmidrule[lr]{3-6}
        \textbf{Photonic nanostructure} & Structure & 2D DBR cavity & DBR + microlens & single-mode WG & nanopillar \\
        ~ & Deterministic alignment & N & N & N & Y \\ \cmidrule[lr]{3-6}
        \textbf{Waveguide coupling} & Type & edge-bonding & photonic wire-bond & chiplet transfer & cryogenic edge-coupling \\
        ~ & Alignment method & RT, passive & RT, imaging & RT, imaging & cryogenic, active \\
        ~ & Permanent & Y & Y & Y & N \\
        ~ & \# WGs coupled & 10 & 9* & 20* & 10 \\
        ~ & SPSs/WG & several & >1 & 10s & 1 \\
        ~ & QD to WG efficiency & 0.08 (sim.) & (0.29 rel.) & 0.83 (sim.) & 0.05 (0.17 rel.) \\
    \bottomrule
    \end{booktabs}
    }
    \caption{\label{tab:table} Comparison of approaches to scalable QD-waveguide integration. DBR: distributed Bragg reflector; WG: waveguide;  RT: room-temperature; *: multi-step process; sim.: simulation of ideal case; rel.: measured QD-waveguide-objective signal relative to QD-objective signal}
\end{table*}

\subsection{Emitter spectral uniformity}

The dominant positive trion line was on average centred at 779 nm (1.59 eV), with a standard deviation of 1.0 nm ($n=39$ QD-pillars), over a region hundreds of microns in dimension. Because of this low inhomogeneous broadening of the system, we observed in several places across the sample that adjacent quantum dots were significantly overlapping in their spectral features. Figure \ref{fig:unif}(a) shows an example of two adjacent QDs that overlapped in the dominant trion line, while (b) shows two different, but similarly overlapping adjacent QDs, simultaneously coupled and observed through adjacent waveguides of the array. In figure \ref{fig:unif}(c) is shown four QDs coupled to the array, with QDs 1 and 3, and 2 and 4 overlapping. We emphasize that in all these cases there were no additional emitters, just a single QD at each array site. This represents a first step towards demonstrating quantum interference between adjacent site-controlled quantum emitters.

\section{Discussion}


Here we have demonstrated for the first time the active alignment and coupling of arrays of site-controlled QDs to arrays of PIC waveguides and devices, marking a significant step for the scalability of quantum photonics. In general, recent high-quality SPSs involve the integration of a simple quantum system, such as a quantum dot or colour centre, within a photonic nanostructure, such as a photonic crystal cavity, bullseye cavity, or micropillar. The engineered photonic environment of the nanostructure enhances the emission quality, rate, and spatial control. Thus, the scalable coupling of SPSs to waveguides will in general involve first the coupling of the emitters to nanostructure modes, and then the coupling of these to the waveguide modes. Here we have demonstrated that both steps can be performed and combined in a scalable way, as the QDs are all deterministically embedded in self-aligned photonic nanostructures, which are then actively aligned to a single-mode waveguide array. Furthermore, the inhomogeneous broadening of the QD distribution remained low, such that adjacent emitters could have a significant spectral overlap without any post-tuning. We emphasize that in this work the site-controlled emitters enabled us to design the PIC in advance, with reference to the known source positions. The alternative, which obviously provides a significant barrier to scalability, is to design the PIC around the random source positions \cite{pfister_telecom_2025}.

As well as the scalability, another advantage of the presented approach is in its chip-to-chip flexibility. In other words, the integration is stable but reversible, and the chips can be separated and realigned, or changed. This enables an easy change of source with respect to a photonic circuit, or moving of sources between different circuits. Here we sequentially couple different rows of site-controlled emitters to the same PIC. Further, the same QD chips were coupled to PICs composed of entirely different material stacks, and different source material architectures were also coupled to the same photonic chip (not shown). These properties may especially be of value in the situation of low device yields, ageing, or damage. The requirements are instead geometrical: in our case the pillar fabrication process results in a planar surface as defined by top of pillars, and thus the entire array can aligned to the plane of the PIC edge facets.

One potential issue is the observed strain. However, even in the case of chip-chip contact, the stress is distributed among the array of nanostructures, and no (permanent) changes or damage were observed to occur in the majority of cases. This contact can be avoided without a significant reduction in the coupling efficiency. With other wider pillar structures \cite{juska_self-aligned_2025} (lateral dimension $>2$ {\textmu}m), the effect of strain was much less, and optimal coupling was achieved with (near-)contact but without any redshift. Other approaches to QD coupling may also have to deal with such strain effects either during the integration process or during cooling of the hybrid system (e.g. a slight blueshift was observed in ref.\cite{pfister_telecom_2025} upon cooling).

Because most other approaches to integration have used randomly positioned QDs, it is difficult to make a straightforward comparison based on the number of QDs coupled, as in general there is a large difference between SPS density and waveguide density. However, we can look at the number of intended SPS-devices coupled, even if in general they contain multiple emitters: see table \ref{tab:table}. Currently, the maximum number of such sources that have been integrated with an array of waveguides in a single step is $\approx 10$, in ref.\cite{ellis_independent_2018}. In that work, a high density of random quantum dots was combined with site-controlled diodes, and several quantum dots were coupled to each waveguide via edge coupling. Here we have matched this using a site-controlled array, in which there is only one QD per grid site, and the yield is close to 100 \%. 

In previous works coupling or integrating multiple single-photon sources, only simulated or relative coupling efficiencies have been able to be reported \cite{murray_quantum_2015,ellis_independent_2018,larocque_tunable_2024,pfister_telecom_2025,wang_large-scale_2025}. For example, in ref. \cite{pfister_telecom_2025}, through photonic wire-bonding, an average waveguide-coupled signal of 0.29 relative to that seen with the objective was obtained, similar to the ratio we have reported above. In ref.\cite{larocque_tunable_2024} 8 micro-chiplets, each consisting of an adiabatic taper and a high density of QDs, were transferred one-by-one in a single fabrication run to a silicon-on-insulator PIC. In that case the theoretical efficiency for an optimally placed QD in an optimally placed chiplet to couple to the Si waveguide was 0.41, although the maximum and average experimental values were not provided. This would have been highly variable due the random positions and orientation of the QDs, with the coupling reducing with both the misorientation of the dipole and the variation in position ($\approx 100$ nm). Similarly, in ref.\cite{wang_large-scale_2025}, 20 GaAs waveguide micro-chiplets were transferred to a lithium niobate PIC. Due to the inclusion of a reflector, the simulated QD-to-waveguide efficiency doubled to 0.83. However, after accounting for the experimental waveguide misalignment range ($\approx 100$ nm lateral, and 7 mrad angular deviation), the maximum coupling (for a perfectly positioned and oriented QD) could reduce to 0.33. The waveguides contained randomly nucleated QDs at a density of 1-10 {\textmu}m$^{-2}$, and the experimental coupling efficiencies across the 20 waveguides were not reported. In all these cases, site-control would be necessary to achieve consistent coupling.

An important parameter for the potential scalability of any coupling/integration strategy is its alignment accuracy relative to its misalignment tolerance. Even with site-control of QDs, the approach will have to be tolerant to the centre-centre variations, in both QD and waveguide position, that will arise due to lithographic limitations. For example, even if electron-beam rather than optical lithography is used to pattern the substrate, uncertainties of 10--50 nm will remain \cite{heindel_quantum_2023}. Here, the nano-positioner resolution is nominally 100 nm over millimetres, and sub-nm over 500 nm at 4 K, and thus the overall alignment accuracy of this approach (for a single QD-waveguide) is sub-nm, an order-of-magnitude better than that currently achievable via other methods \cite{heindel_quantum_2023}. For coupling multiple QD-waveguides simultaneously, the biggest factor is therefore the centre-centre variations. As mentioned above, the QD positions vary by less than 50 nm from the ideal lattice node, and the waveguides by less than 25 nm. Our nanopillars theoretically achieve up to 90 \% of their maximum coupling at displacements up to 100 nm. Thus, the tolerance of the presented approach was not limited by lithographic or alignment constraints, but rather the edge-coupling efficiency of the fabricated photonic nanostructures.

\subsection{Improvements and future directions}

Improvements to the waveguide-coupling efficiency may be achieved by tailoring the waveguide facets and pillar structures further, for improved directionality and mode-matching. The performance of the fabricated nanopillars were a limiting factor, and these operated close to their simulated capability. In the simulations an important variable was the QD to pillar top distance \cite{colavecchi_optimizing_2025}. Additionally, simulations and other experimental results on other samples, suggest that narrower pillars are in general brighter, and that tapering the pillars may allow for better mode-matching to the waveguide.

In our approach, the coupling of all sources and waveguides is performed simultaneously, and thus the coupling of, e.g., 1000 QDs will not entail significantly more resource than two, in contrast with one-by-one processes (e.g. one at a time photonic wire-bonding or micro-chiplet transfer). We have verified the simultaneous coupling of $\approx 10$ QDs, but in principle hundreds are possible with the described setup and procedure. To support this, we note that, while only the waveguide array was intentionally matched to the QD array, when this was aligned, the MMI, which was 600 {\textmu}m away, was still partially coupled (see SI). Only a small adjustment along the array direction was required, indicating that it is possible to couple devices at least this far apart by matching all of the PIC devices to the QD grid. However, in this case, if GaAs and SiN chips are again used, the differential contraction upon cooling would have to be accounted for. Over the 100 {\textmu}m wide array described above, the calculated difference is approximately 50 nm, and was not observed to affect the coupling (in agreement with simulations). Over 1 cm however, this difference would become 5 {\textmu}m, and thus a correction factor would need to be applied to the waveguide positions during the design of the PIC.

Denser integration could be also achieved by stacking waveguides or coupling to laser-written 3D circuits \cite{zhong_laser-direct-lithography_2025}, an approach not obviously compatible with current transfer printing or pick-and-place schemes. This will increase the challenge of exciting multiple QDs simultaneously. For this purpose, ring resonators can be used to couple to multiple QDs, and to demultiplex the laser from the QD signals. If the QDs are grown in a diode structure, electrical excitation also becomes an option.

In terms of the alignment procedure and experimental arrangement, rotation piezoelectric stages would aid in ensuring that the coupling across the chip is uniform. Fiber-coupling could also be used at the detector side of the PIC \cite{ellis_independent_2018, psiquantum_team_manufacturable_2025} to improve the collection efficiency and the number of PIC device outputs simultaneously accessible. With modifications to the QD stack, room temperature alignment and bonding could also become feasible.

\section{Conclusion}

Significant steps towards a scalable quantum photonic computing architecture were demonstrated through the active alignment of arrays of site-controlled deterministic quantum light sources to arrays of nanophotonic waveguides. The alignment was flexible and reconfigurable at cryogenic temperatures, while maintaining stability. We verified the high single-photon purity and the low inhomogeneous broadening of the site-controlled emitter array. Future improvements in the SPS-nanostructure quality and in the coupling efficiency can enable scalable demonstrations of fundamental quantum information protocols.

\begin{acknowledgments}
This work has received funding from the Enterprise Ireland's DTIF programme project No. DT 2019 0090B. We also acknowledge the University of Southampton and the UK Engineering and Physical Sciences Research Council (EPSRC) CORNERSTONE 2 project EP/T019697/1 for providing fabrication of some of the devices in this work. GJ and EP also acknowledge funding from Research Ireland, formerly Science Foundation Ireland, under Grants Nos. 22/FFP-P/11530, 22/FFP-A/10930, 15/IA/286
\end{acknowledgments}

\section{Author Contributions}

JO'H, NM, and MJ developed the coupling scheme. JO’H performed the coupling experiments and correlation measurements, calculations and modelling. NM designed the cryogenic mount and performed the simulations of the pillar coupling. NM and SM fabricated the pillar with contribution from MJ. MJ managed the back-end processing of the SiN chips. JH designed the SiN chips with inputs from NM, MJ and JO'H. JH performed the room temperature characterization of the PICs. SM designed and fabricated PICs for testing alternative platforms. GJ performed optical characterization of the QD samples and correlation measurements. LC grew the epitaxial structure of the QD samples. FHP, BC, and EP initiated the project and provided expertise and oversight. JO’H wrote the paper with contributions from all authors.

\section{Competing interests}

The authors declare no competing interests.

\section{Materials \& Correspondence}

Correspondence and material requests should be addressed to john.ohara@tyndall.ie.

\section{Methods}

\subsection{Spectra acquisition and filtering}

In all cases the signal from the QD or waveguide was sent to a monochromator with a 950 lines/mm grating. In the case of above-band excitation (636 nm), a 700 nm longpass filter was used to filter the laser before the monochromator; and in the case of the quasi-resonant excitation, between one and three notch filter were placed in the collection path(s) in order to suppress the laser. For the spectra, the signal dispersed by the grating was collected by a silicon CCD. For the single-photon measurements, the signal was sent to the exit slit of the monochromator, which was fixed at 0.3 mm, corresponding to a 0.2 nm filtering window around the line of interest. 

\subsection{Correlation measurements}

For the $G^{(2)}(\tau)$ measurements, the two signals were sent through two separate monochromators, and registered by avalanche photo-diodes. These were correlated with a single-photon counter module. 50 ps bins were used to obtain the raw data, and these were converted to 2.5 ns bins for display and analysis. The $g^{(2)}_{\text{raw}}(0)$ was calculated by summing the three central bins (7.5 ns window) around each peak (every 12.5 ns) and dividing the value at zero time delay by the average value in the equivalent windows for the side peaks. No background subtraction was performed. The standard deviation of the values for the side peaks was approximately equal to the Poissonian noise $\sqrt {N}$, and by assuming the zero-delay peak was also affected by Poissonian noise, the uncertainty in the $g^{(2)}_{\text{raw}}(0)$ was then obtained via a min-max approach to the peak values.

\subsection{Array ratio calculation}

The ratio of the signals observed through the waveguide and directly with the objective were determined by performing Voigt peak fits of the dominant lines in the spectra. Different fitting procedures did not differ from this value (0.17) by more than 0.01, and since the linewidths were all similar during and after coupling, comparison of the peak CCD counts alone also gave a similar mean value (0.16). The uncertainty was obtained via a min-max approach.

\subsection{Tabletop efficiency}

The power of a 780 nm laser was measured before entering the back of the objective, as well as after the multi-mode fiber which was connected to the single-photon detector. Correcting for the losses due to overfilling and passing through the objective, along with an extra reflection and an extra pass through the cryostat window, and using the specification of the detector at 780 nm, we obtained an overall efficiency of $0.064 \pm 0.005$ at 780 nm. This was reduced to 0.056 when using a notch filter, and 0.041 when using three notch filters and a beam-splitter. More details are provided in the SI.

\subsection{Temperature monitoring}

The temperature of the QD sample was monitored using a sensor on top of the $xyz$ stage, while the temperature of the SiN chip was monitored using the temperature-dependent PL of a triple GaAs quantum well in close proximity on its copper mount.


\end{document}